\documentclass[conference]{IEEEtran}
\IEEEoverridecommandlockouts
\usepackage{cite}
\usepackage{amsmath,amssymb,amsfonts}
\usepackage{algorithmic}
\usepackage{graphicx}
\usepackage{textcomp}
\usepackage{xcolor}

\def\BibTeX{{\rm B\kern-.05em{\sc i\kern-.025em b}\kern-.08em
    T\kern-.1667em\lower.7ex\hbox{E}\kern-.125emX}}
\begin{document}

\title{
Multiplexed Control at Scale for Electrode Arrays in Trapped-Ion Quantum Processors
\thanks{
This work was supported by MEXT Q-LEAP (Grant Number JPMXS0120319794), JST (Grant Number JPMJPF2014), and JST Moonshot R\&D (Grant Numbers JPMJMS226A, JPMJMS2063).
\\© 2025 IEEE.  Personal use of this material is permitted.  Permission from IEEE must be obtained for all other uses, in any current or future media, including reprinting/republishing this material for advertising or promotional purposes, creating new collective works, for resale or redistribution to servers or lists, or reuse of any copyrighted component of this work in other works.
}
}

\author{
    \IEEEauthorblockN{
        Ryutaro~Ohira\IEEEauthorrefmark{1},
        Shinichi~Morisaka\IEEEauthorrefmark{1}\IEEEauthorrefmark{2}, 
        Ippei~Nakamura\IEEEauthorrefmark{3}, Atsushi~Noguchi\IEEEauthorrefmark{3}\IEEEauthorrefmark{4}\IEEEauthorrefmark{5}, 
        and Takefumi~Miyoshi\IEEEauthorrefmark{1}\IEEEauthorrefmark{2}\IEEEauthorrefmark{6}
    }
    \IEEEauthorblockA{\IEEEauthorrefmark{1}QuEL, Inc., Daiwaunyu Building 3F, 2-9-2 Owadamachi, Hachioji, Tokyo, Japan}
    \IEEEauthorblockA{\IEEEauthorrefmark{2}Center for Quantum Information and Quantum Biology, Osaka University, 1-2 Machikaneyama, Toyonaka, Osaka, Japan}
    \IEEEauthorblockA{\IEEEauthorrefmark{3}Komaba Institute for Science (KIS), The University of Tokyo, 3-8-1 Komaba, Meguro, 153-8902, Tokyo, Japan}
    \IEEEauthorblockA{\IEEEauthorrefmark{4}RIKEN Center for Quantum Computing (RQC), 2-1 Hirosawa, Wako, 351-0198, Saitama, Japan}
    \IEEEauthorblockA{\IEEEauthorrefmark{5}Inamori Research Institute for Science (InaRIS), 620 Suiginya, Kyoto, 600-8411, Kyoto, Japan}
    \IEEEauthorblockA{\IEEEauthorrefmark{6}e-trees. Japan, Inc., Daiwaunyu Building 2F, 2-9-2 Owadamachi, Hachioji, Tokyo, Japan}

    ohira@quel-inc.com, morisaka@quel-inc.com, ippei-nakamura@g.ecc.u-tokyo.ac.jp, \\u-atsushi@g.ecc.u-tokyo.ac.jp, miyoshi@quel-inc.com
}

\maketitle

\begin{abstract}
The scaling up of trapped-ion quantum processors based on the quantum charge-coupled device (QCCD) architecture is difficult owing to the extensive electronics and high-density wiring required to control numerous trap electrodes. 
In conventional QCCD architectures, each trap electrode is controlled via a dedicated digital-to-analog converter (DAC).
The conventional approach places an overwhelming demand on electronic resources and wiring complexity.
This is because the number of trap electrodes typically exceeds the number of trapped-ion qubits.
This study proposes a method that leverages a high-speed DAC to generate time-division multiplexed signals to control a large-scale QCCD trapped-ion quantum processor. 
The proposed method replaces conventional DACs with a single high-speed DAC that generates the complete voltage waveforms required to control the trap electrodes, thereby significantly reducing the wiring complexity and overall resource requirements.
Based on realistic parameters and commercially available electronics, our analysis demonstrates that a QCCD trapped-ion quantum computer with 10,000 trap electrodes can be controlled using only 13 field-programmable gate arrays and 104 high-speed DACs.
This is in stark contrast to the 10,000 dedicated DACs required by conventional control methods. 
Consequently, employing this approach, we developed a proof-of-concept electronic system and evaluated its analog output performance.
\end{abstract}


\section{Introduction}

Recent developments that incorporate tens of trapped-ion qubits~\cite{pino2021demonstration, moses2023race, decross2024computational} have demonstrated that the quantum charge-coupled device (QCCD) architecture offers a route for scalable trapped-ion quantum computers~\cite{wineland1998experimental, kielpinski2002architecture, lekitsch2017blueprint, loschnauer2024scalable}. 
In the QCCD architecture, trapped-ion qubits are physically shuttled between distinct functional areas, such as state preparation, measurement, and gate operations, to perform quantum computations.

In a typical trapped-ion QCCD system, ion shuttling is achieved by dynamically controlling the trap potentials via time-varying voltages applied to the trap electrodes. 
Conventionally, each electrode is controlled by a dedicated digital-to-analog converter (DAC) operating at voltage update rates ranging from several hundred kHz to a few MHz~\cite{blakestad2010transport, blakestad2011near, kaushal2020shuttling, akhtar2023high, mordini2025multizone}. 

However, this approach poses significant challenges in terms of scaling up the system. 
In a typical QCCD architecture, a single trapped-ion qubit requires multiple trap electrodes, with current estimates indicating the need for approximately 10 electrodes per qubit~\cite{pino2021demonstration, kaushal2020shuttling, malinowski2023wire}. 
This one-to-one correspondence between the DACs and trap electrodes increases wiring complexity that scales approximately tenfold with the number of qubits.
These challenges are further exacerbated by the need to interface vacuum or cryogenic systems with room-temperature control electronics.
Several strategies have been proposed to address these issues, including the use of in-vacuum~\cite{guise2014vacuum} or chip-integrated~\cite{stuart2019chip} electronics, switching networks~\cite{malinowski2023wire}, electrode co-wiring techniques~\cite{moses2023race}, and novel transport operation methods that reduce the number of required analog signals~\cite{delaney2024scalable}.

\begin{figure*}[t]
\begin{center}
\includegraphics[width=.96\textwidth]{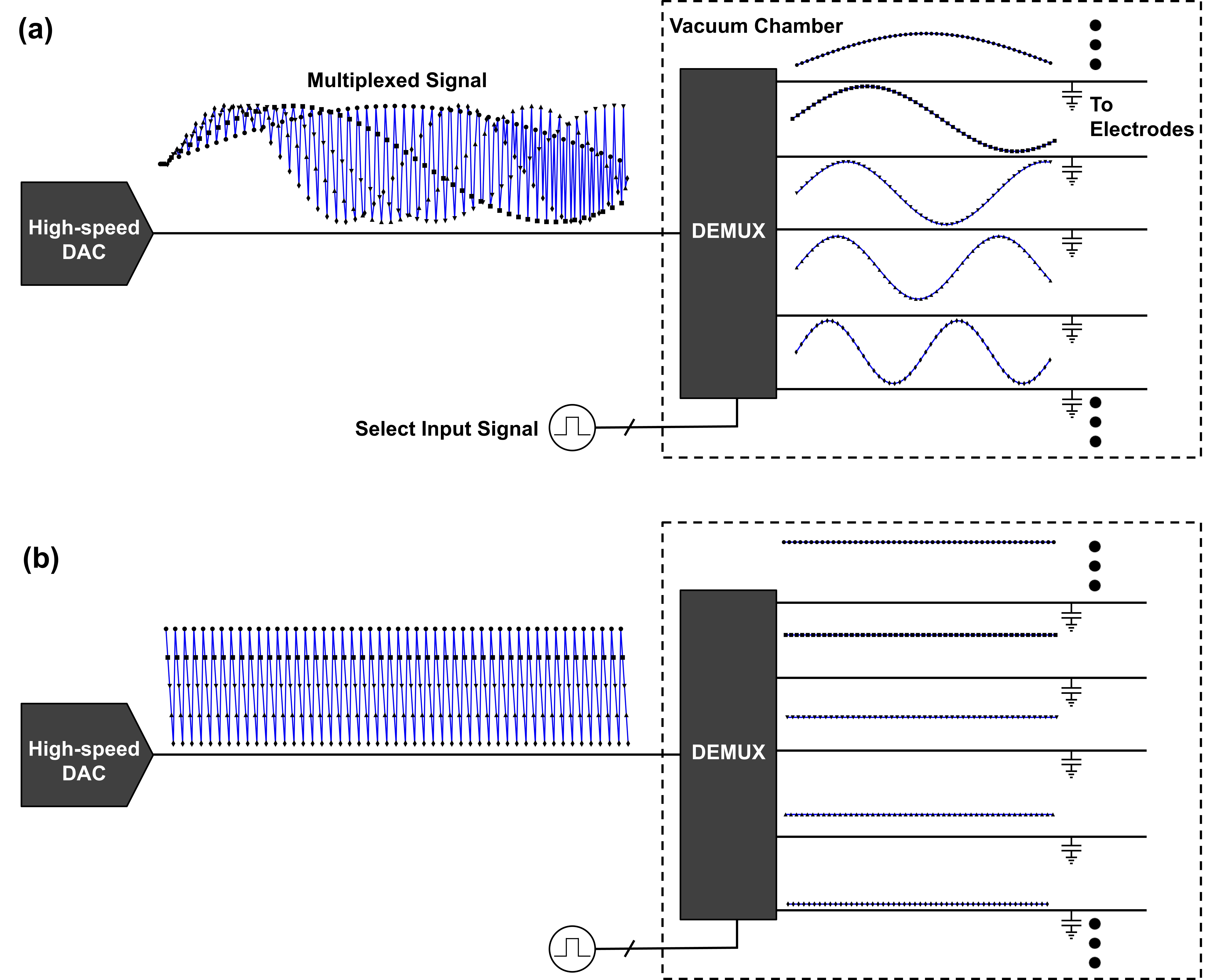}
\end{center}
\caption{
Proposed electrode control scheme. 
The proposed method leverages a high-speed digital-to-analog converter (DAC) that operates at sampling rates considering exceeding those of conventional DACs typically used for ion shuttling operations. 
The high-speed DAC facilitates time-division multiplexing across multiple channels by generating complete voltage waveforms, thereby eliminating the requirement of a dedicated DAC for each electrode. 
The generated waveform is then routed through a demultiplexing network (denoted as DEMUX), which distributes the signal to each output channel, where a capacitor holds the voltage until the next update cycle. 
(a) Time-multiplexed signal composed of sinusoidal waves at different frequencies. 
This waveform demonstrates that our system can generate the time-dependent signals required for the dynamic control of the trapped-ion qubits. 
(b) Static voltage signal suitable for ion trapping.
}
\label{fig:general_idea}
\end{figure*}

\begin{figure*}[t]
\begin{center}
\includegraphics[width=.96\textwidth]{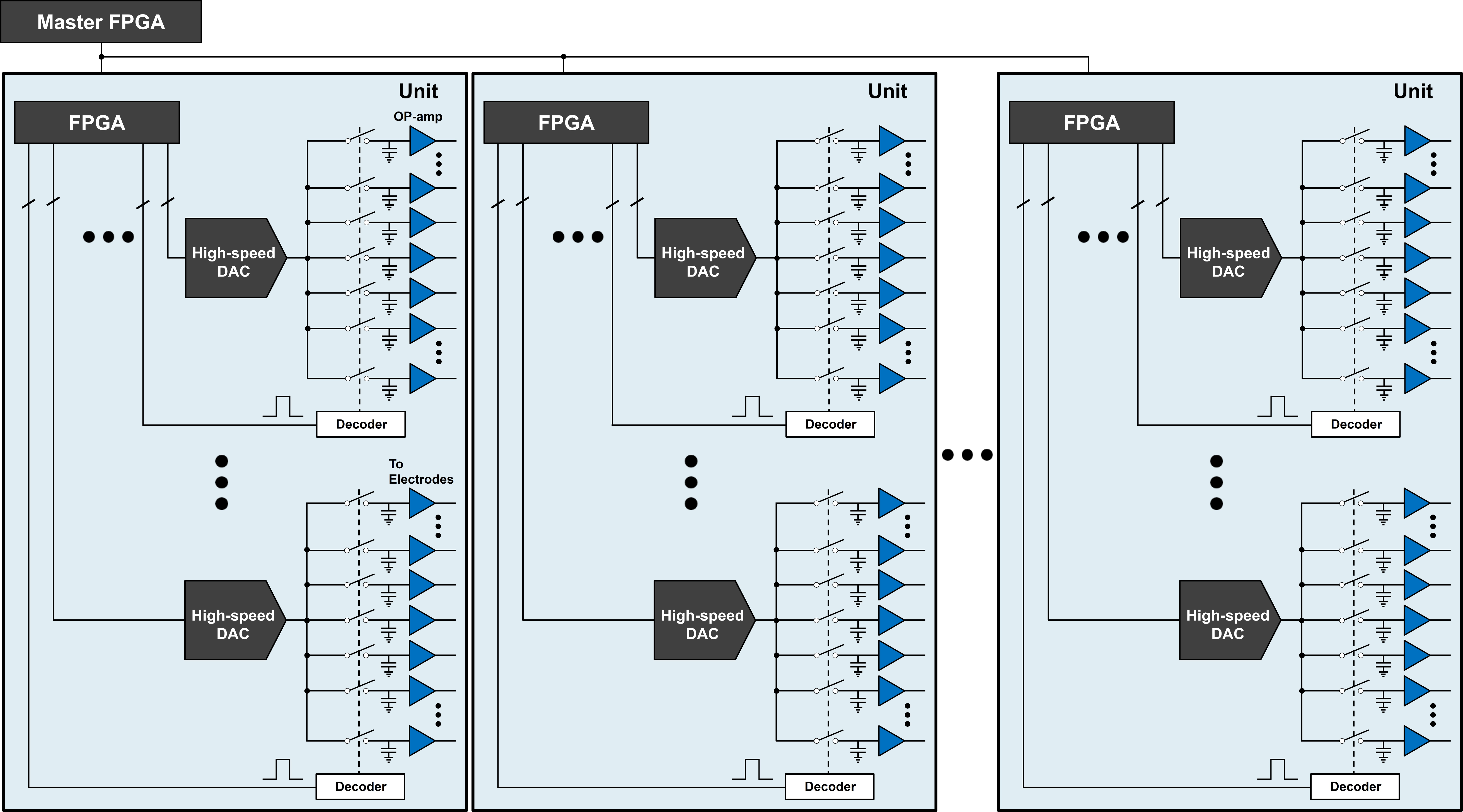}
\end{center}
\caption{
Practical implementation of the proposed scheme. 
High-speed switches and a digital decoder (denoted as Decoder) sequentially charge the capacitors positioned before each electrode. 
Upon charging, the operational amplifier (denoted as OP-amp) boosts the voltage to the desired level. 
The limited number of field-programmable gate arrays (FPGA) input and output (I/O) ports restricts the number of DACs that can be controlled and the available digital signals for demultiplexing. 
Thus, each control unit is designed as a modular element, and synchronization signals are distributed among them to enable the modules to operate together as a unified control system. 
In our implementation, a master FPGA coordinates all the control units, thereby ensuring synchronized operation under a common clock signal.
}
\label{fig:implementation}
\end{figure*}

This study proposes an alternative strategy to decrease both the number of DACs and the wiring complexity in large-scale QCCD trapped-ion quantum processors.
The proposed method employs a high-speed DAC that has been configured to generate time-division multiplexed signals capable of replicating the outputs of multiple conventional DACs. 
Subsequently, these signals are distributed across individual channels to generate the voltage waveforms required to control the trap electrodes. 

Our analysis demonstrates that a QCCD quantum computer with 10,000 trap electrodes can be controlled with only 13 field-programmable gate arrays (FPGAs) and 104 high-speed DACs, based on realistic parameters and commercially available components. 
This significantly reduces hardware demands compared to 10,000 dedicated DACs required by conventional control methods.
Furthermore, we developed a proof-of-concept (PoC) electronic system and evaluated the quality of its analog output signals.

\section{
General Concept
}\label{sec:general_concept}

This section elucidates the proposed control scheme and describes a detailed implementation of the proposed approach. 
Notably this study develops a control method for trapping electrodes within a large-scale QCCD architecture. 
The precision of this approach in terms of controlling the quantum states of trapped-ion qubits will be examined in future research. 
In addition, shuttling ions on a timescale shorter than their motional period requires the use of DACs with voltage update rates on the order of tens of MHz~\cite{walther2012controlling, bowler2012coherent, bowler2013arbitrary, baig2013scalable, sterk2022closed, clark2023characterization}. 
However, this study focuses on adiabatic transport and does not consider such high-speed ion shuttling operations.

\subsection{
Proposed System Architecture
}

Figures~\ref{fig:general_idea}(a) and (b) present an overview of the proposed method. 
The proposed method is based on a high-speed DAC that operates at a sampling rate significantly higher than several mega samples per second (Msps) typically used for ion shuttling. 
This high-speed DAC generates the complete voltage waveforms required to control all the responsible electrodes, enabling time-division multiplexing across multiple channels. 
The output of the DAC is routed through a demultiplexing network, which distributes the generated waveforms to individual electrode channels. 
Each channel includes a capacitor that is charged to maintain the desired voltage.

Figure~\ref{fig:general_idea}(a) shows a time-multiplexed signal composed of sinusoidal waves at different frequencies. 
Figure~\ref{fig:general_idea}(a) serves as an illustrative example, demonstrating that our scheme can encode arbitrary time-dependent signals required for the dynamic control of trapped-ion qubits. 
In contrast, when ions are held statically in a trap, a signal like that depicted in Fig.~\ref{fig:general_idea}(b) is generated.
Note that in a typical trapped-ion setup, the DAC output must be low-pass filtered to suppress the electrical noise that could excite the ions' motional state. 
For simplicity, the low-pass filters are omitted from the depiction in this study.

In principle, the proposed control method can also be applied to static ion trapping. 
However, as described later, the voltage drop and recharging cycle may induce the motional excitation of the ions or affect the stability of their motional frequencies during gate operations, thereby degrading the quantum gate fidelity~\cite{malinowski2023wire}. 
These issues warrant further theoretical and experimental investigation.

Figure~\ref{fig:implementation} illustrates a practical implementation of the proposed scheme. 
An FPGA generates a time-multiplexed signal and then sends it to a high-speed DAC. 
The time-multiplexed signal is demultiplexed employing high-speed switches and a digital decoder (denoted as Decoder in Fig.~\ref{fig:implementation}). 
Each high-speed switch is activated at the appropriate moment to charge the associated capacitor. 
Upon the charging of the capacitor, the switch turns off, and the next switch is turned on to charge the adjacent capacitor. 
Subsequently, the voltage stored in each capacitor is amplified via an operational amplifier (OP-amp) to achieve the desired level.

This operational principle is partially inspired by the voltage distribution scheme for static ion control, as proposed by Malinowski et al.~\cite{malinowski2023wire}. 
However, the proposed approach is distinctly different from that of Malinowski et al.~\cite{malinowski2023wire}.
In their scheme, the ion qubits are dynamically controlled through the combination of DAC outputs that operate at update rates of a few MHz with an integrated switch network. 
Meanwhile, the quasi-static control employs a sample-and-hold circuit and a demultiplexer to control the electrodes using a limited number of DACs. 
In contrast, the proposed approach consolidates both dynamic and quasi-static control using a small number of high-speed DACs to directly generate all necessary voltage waveforms, which are then demultiplexed to individually address the electrodes.

Both the number of DACs that can be controlled and the number of digital signals that can be sent to the digital decoder are restricted due to the limited number of input and output (I/O) ports on an FPGA. 
To address this issue, each control unit is designed as a modular element, as illustrated in Fig.~\ref{fig:implementation}, and synchronization signals are distributed among them, enabling the units to collectively function as a single control system. 
Synchronization between individual control units can be achieved using methods such as those described in Refs.~\cite{miyoshi2022fpl, miyoshi2024microwave, miyoshi2025toward}. 
For example, all control units can be connected to a master FPGA via 10 GbE and controlled using a shared common clock~\cite{miyoshi2022fpl, miyoshi2024microwave, miyoshi2025toward}.

\begin{figure}[t]
\begin{center}
\includegraphics[width=.45\textwidth]{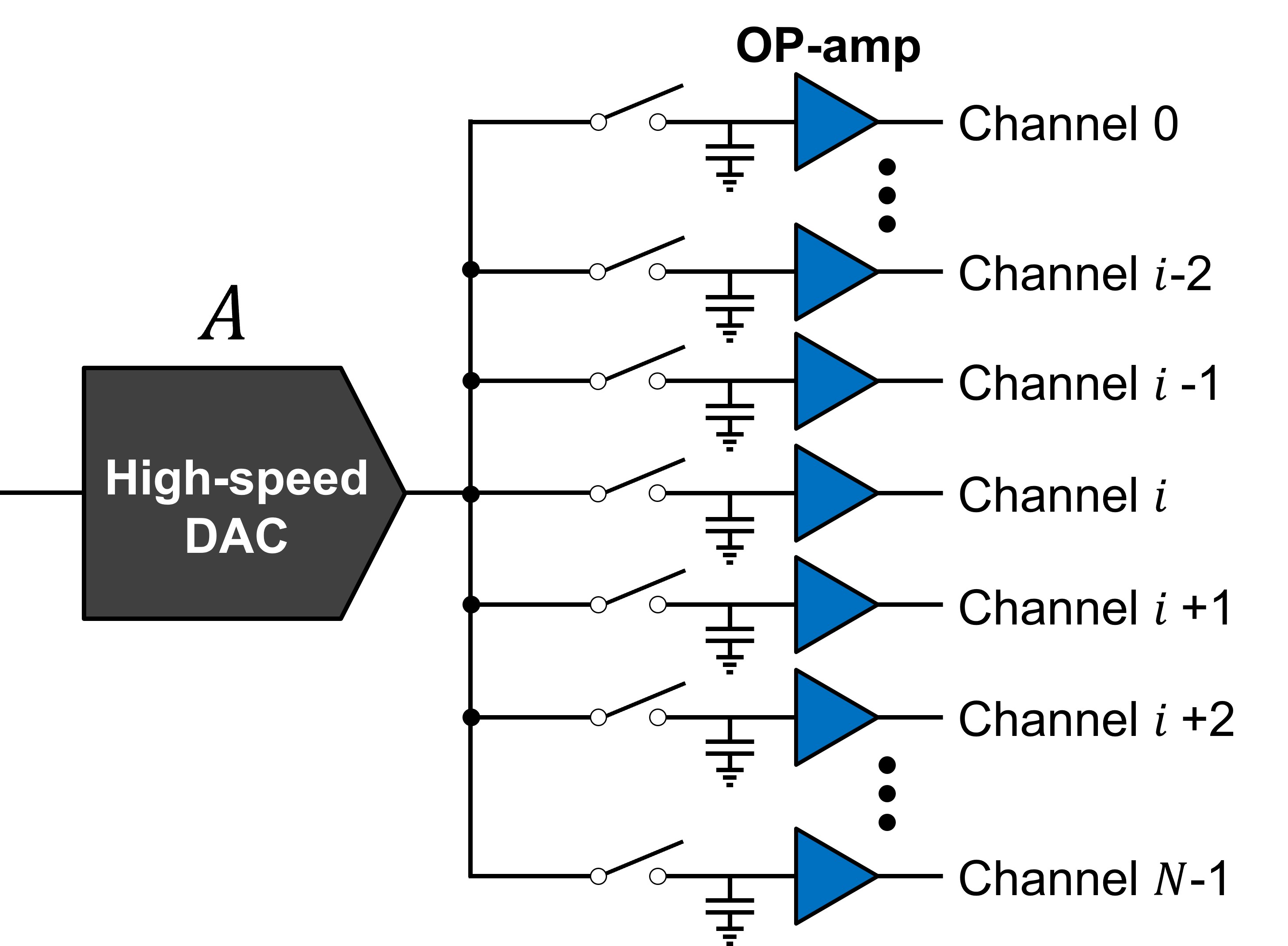}
\end{center}
\caption{
Multiplexing factor is the number of electrodes that are controlled using a single high-speed DAC. 
For example, if a high-speed DAC operating at $A$ distributes voltage to $N$ channels, the multiplexing factor is $N$.
}
\label{fig:multiplexing_factor}
\end{figure}

\begin{figure*}[t]
\begin{center}
\includegraphics[width=.95\textwidth]{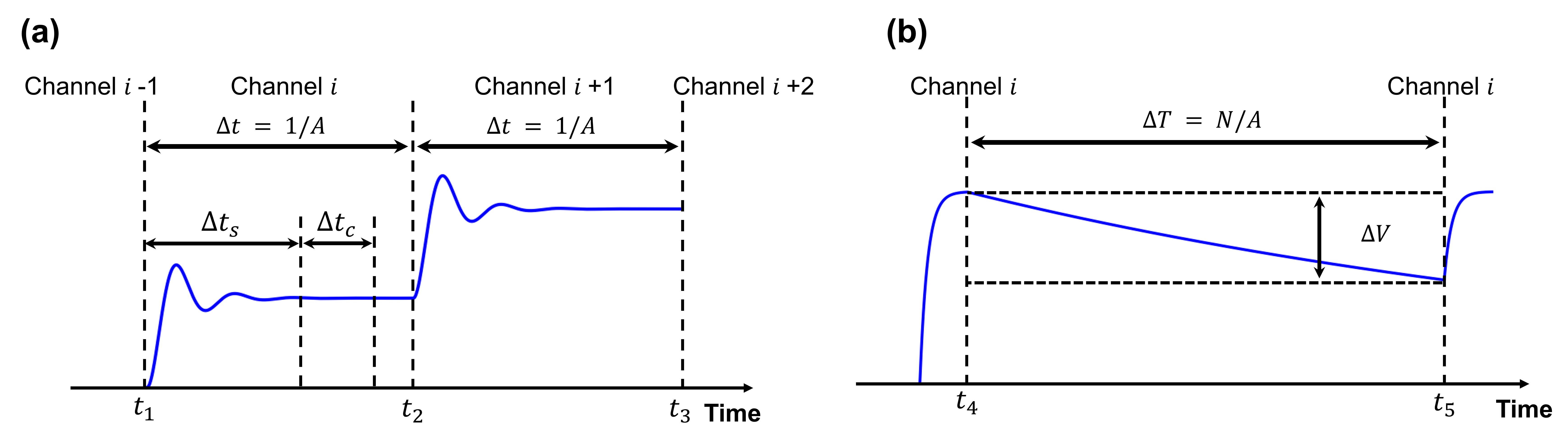}
\end{center}
\caption{
(a) The vertical and horizontal axes show the voltage delivered to each channel's capacitor and time (marked at $t_1$, $t_2$, and $t_3$), respectively. 
At $t_1$, the system switches from Channel $i-1$ to Channel $i$.
This causes the DAC output to transition from the target voltage of Channel $i-1$ to that of Channel $i$. 
The high-speed DAC output necessitates a settling time $\Delta t_s$ for stabilization at the new target voltage, followed by a capacitor charging time $\Delta t_c$. 
Assuming the sum of the settling time $\Delta t_s$ and capacitor charging time $\Delta t_c$ to be less than the total time available per channel $\Delta t$, the capacitor reaches the desired voltage.
(b) Voltage profile of the capacitor in Channel $i$ over time.
At $t_4$, the capacitor in Channel $i$ reaches full charge; however, its voltage begins to decay until the next charging event at $t_5$.
The duration between these charging events is $\Delta T =N\Delta t = N/A$.
}
\label{fig:limiting_factor}
\end{figure*}

\subsection{
Estimation of Multiplexing Factor of Proposed Method
}

Now, we estimate the achievable multiplexing factor and overall scalability of our the proposed method. 
Here, the multiplexing factor is the number of electrodes that can be controlled using a single high-speed DAC. 
Figure~\ref{fig:multiplexing_factor} shows a schematic of a high-speed DAC operating at rate $A$ distributing voltage to $N$ channels (Channel 0 to $N$-1). 
In this configuration, the multiplexing factor is defined as $N$.

Two primary factors limit this multiplexing factor. 
(1) The time required to fully charge the capacitor. 
(2) The voltage drop owing to the capacitor discharge. 
Here, we examine the impact of each factor on the achievable multiplexing factor.

\subsubsection{
Capacitor charging time
}

First, we describe the effect of the time required to charge the capacitor on the achievable multiplexing factor. 
Consider a situation wherein a high-speed DAC operating at $A$ distributes voltage to $N$ channels (Channel 0 to $N$-1), as illustrated in Fig.~\ref{fig:multiplexing_factor}. 
At this operating rate, the time allocated to each channel is 
\[
\Delta t = 1/A.
\]
Within this interval, the capacitor associated with each channel must be completely charged.

Two key parameters affect this process: 
\begin{itemize}
    \item The settling time of the high-speed DAC
    \item The charging time of the capacitor.
\end{itemize} 
The settling time is the duration between the setting of an input code and the output stabilization at its final value. 
If charging begins prior to the settling of the output of the high-speed DAC, the capacitor may reach an incorrect voltage. 
Therefore, the high-speed DAC output must first be allowed to stabilize before turning on the switch to charge the capacitor. 
We denote the high-speed DAC settling time as $\Delta t_s$ and the capacitor charging time as $\Delta t_c$. 
Consequently, the sum $\Delta t_s + \Delta t_c$ must be less than the available time per channel, $\Delta t$.

Figure~\ref{fig:limiting_factor}(a) illustrates the effects. 
The vertical and horizontal axes represent the voltage delivered to the capacitors of each channel and time (marked at specific instants $t_1$, $t_2$, and $t_3$), respectively. 
At time $t_1$, the system switches from Channel $i-1$ to $i$. 
This causes the high-speed DAC output to transition from the target voltage for Channel $i-1$ to that of Channel $i$. 
The high-speed DAC output requires a settling time $\Delta t_s$ to stabilize at the new target voltage, followed by an additional time $\Delta t_c$ for the capacitor to fully charge. If
\[
\Delta t_s + \Delta t_c < \Delta t,
\] 
the capacitor can achieve the desired voltage.

This combined time, $\Delta t_s + \Delta t_c$, effectively limits the maximum sampling rate of the high-speed DAC that can be used with this control scheme. 
A high-speed DAC must operate at a rate no greater than 
\[
\frac{1}{\Delta t_s + \Delta t_c}
\]
to ensure proper charging of the capacitor. 
Notably, the finite rise and fall times of the switches also affect the capacitor charging time $\Delta t_c$. 
In practice, $\Delta t_c$ is determined by the sum of the switching times (both turn-on and turn-off) of the switch and the on duration of the switch.

\subsubsection{
Capacitor discharge-induced voltage drop
}

We examine the impact of the voltage drop due to the capacitor discharge between successive charging events. 
The voltage evolution of the capacitor in Channel $i$ over time is schematically shown in Fig.~\ref{fig:limiting_factor}(b).
At time $t_4$, the capacitor in Channel $i$ is fully charged. 
Over time, the voltage decreases until the next charging event occurs at time $t_5$. 
The interval between the completion of one charging cycle and the subsequent recharge is given by
\[
\Delta T = N\Delta t = N/A.
\]

To minimize the voltage variations from the capacitor discharge, the discharge time constant should be designed to be as long as possible. 
One straightforward approach is to use a capacitor with a larger capacitance. 
However, increasing the capacitance also prolongs the charging time. 
Furthermore, if the multiplexing factor $N$ is too high, the charging cycle becomes longer, leading to a more pronounced voltage drop.

After identifying the factors limiting the multiplexing factor, we then estimate it by defining the target signal quality. 
In this study, we consider multiplexing a DAC with a 0.5~MHz voltage update rate. 
To establish concrete circuit parameters, we assume the following:
\begin{itemize}
    \item A high-speed DAC with a settling time of $\Delta t_s=10$~ns.
    \item High-speed switches with a rise/fall time of $\tau_{\rm sw}=1$~ns and an on-resistance of $R_{\rm sw}=10~\Omega$.
    \item A $C=150$~pF capacitor for each channel.
    \item An OP-amp with suitable gain and an input impedance of $R_{\rm amp}=10~\rm M\Omega$.
\end{itemize}

First, the capacitor charging time is determined based on the RC time constant as $\tau_c=CR_{\rm sw}=1.5$~ns.
Typically, approximately 5 times this RC constant is required by the capacitor to charge nearly to its final value. 
Thus, the charging time is approximately $5\tau_c=7.5$~ns.
Including the 1-ns rise and 1-ns fall times of the switch, we set the effective capacitor charging time to $\Delta t_c=1+7.5+1=9.5$~ns.
Thus, the total time required to charge the capacitor for each channel is $\Delta t_s+\Delta t_c=10+9.5=19.5$~ns.
This corresponds to the maximum sampling rate for high-speed DAC of approximately $1/(\Delta t_s+\Delta t_c) \approx51.3$~Msps.
Assuming the high-speed DAC operates at a sampling rate of $A=50~\rm Msps$ and targeting a voltage update rate of 0.5~MHz for individual output channels, the multiplexing factor is obtained as $N=50/0.5=100$.

Meanwhile, the time interval between successive charging events for a particular channel is $\Delta T=N/A=2~\rm\mu$s.
The discharge time constant of the capacitor is $\tau_d=CR_{\rm amp}=1.5$~ms.
Using the exponential decay model $V_0 e^{-t/\tau_d}$, after 2~$\rm\mu$s the voltage decreases by approximately 0.13\% relative to the initial voltage $V_0$.

Several strategies can be employed to further enhance the multiplexing factor. 
The above estimates are based on the performance of off-the-shelf electronics. 
If higher-performance electrical components are used, further improvements in multiplexing capability can be achieved. 
For example, from the viewpoint of high-speed DACs, using a DAC with a shorter settling time would enable faster operation. 
In addition, from the viewpoint of capacitor charging, replacing the switch with a lower on-resistance would allow for even faster charging.

\begin{figure*}[t]
\begin{center}
\includegraphics[width=.95\textwidth]{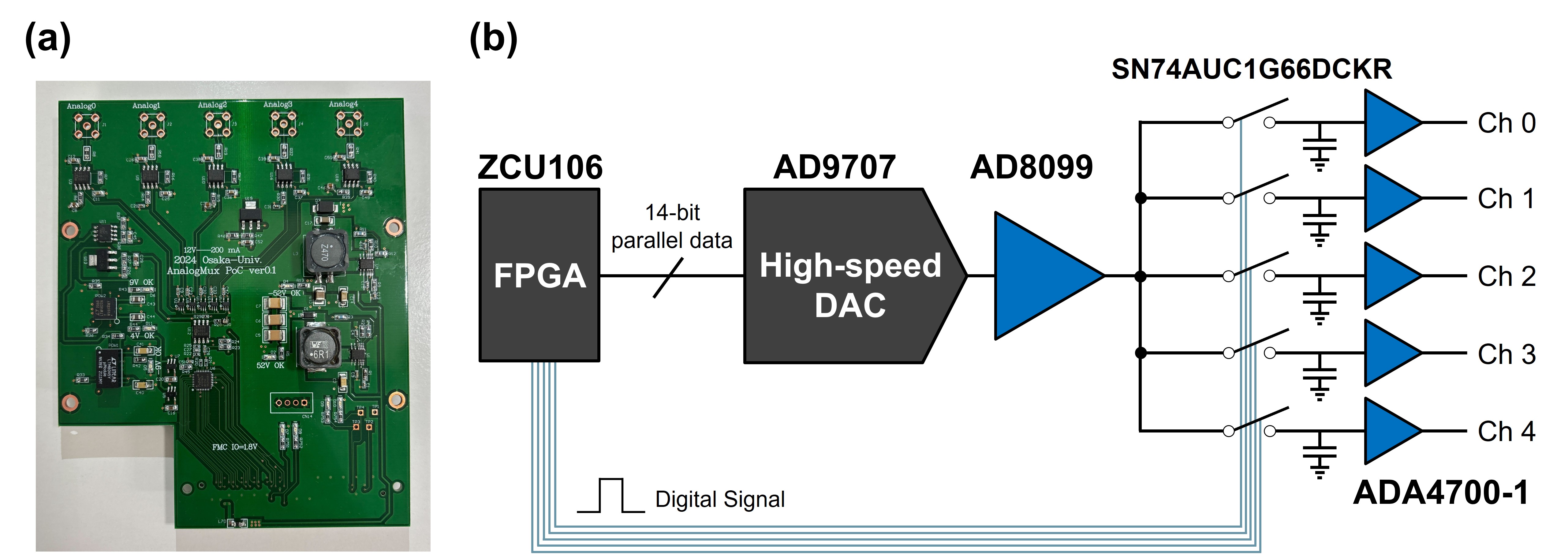}
\end{center}
\caption{
(a) Developed proof-of-concept (PoC) System. 
The FPGA used to control the high-speed DAC is not shown herein. 
(b) Developed system used to multiplex five output signals. 
The design uses a 14-bit high-speed DAC (AD9707) with amplified output (AD8099) distributed to each channel. 
Each channel includes a high-speed switch (SN74AUC1G66DCKR), a 30-pF capacitor, and an OP-amp (ADA4700-1) that produces signals from --50 to +50~V. 
Note that for this PoC, the board does not include a dedicated digital decoder. 
Here, demultiplexing is achieved by directly routing digital signals to each switch. 
The high-speed DAC is controlled using the AMD ZCU106 Evaluation Board employing a 14-bit parallel interface. 
Further, the FPGA also generates the digital control signals for the switches.
}
\label{fig:design}
\end{figure*}

\subsection{
Scalability
}

Finally, we examine the scalability of the proposed control method, considering the limited number of FPGA I/O ports. 
As a target scenario, we consider controlling 10,000 electrodes (corresponding to 1,000 qubits, assuming 10 electrodes per qubit). 
Suppose we use a typical FPGA with 200 I/O ports; however, FPGAs with more than 200 I/O ports are commonly available for IO-intensive applications. 
These I/O ports must support both the interface between the FPGA and the DAC and the digital signals sent to the digital decoder, where the number of decoder signals scales as \(\lceil \log_2 N \rceil\) for $N$ channels.

We consider the usage of a 16-bit parallel data transfer interface for the high-speed DAC, which requires 16 data lines. 
In addition, one I/O port is required for the FPGA-provided clock signal, resulting in 17 I/O ports for the FPGA-DAC interface. 
With a multiplexing factor of 100 (i.e., one high-speed DAC controls 100 electrodes), the digital decoder requires approximately \(\lceil \log_2 100 \rceil \approx 7\) FPGA I/O ports. 
Thus, each high-speed DAC and decoder module uses 24 FPGA I/O ports.

Given an FPGA with 200 I/O ports, one FPGA can support approximately 8 modules. 
Since each module controls 100 electrodes, one FPGA can manage approximately $8 \times 100 = 800$ electrodes.
Therefore, controlling 10,000 electrodes would require \(\lceil 10,000/800 \rceil \approx 13\) FPGAs.
The total number of high-speed DACs needed is $13\times8=104$.
Note that synchronization signals between control units are necessary, as shown in Fig.~\ref{fig:implementation}, these are provided via the high-speed transceivers of the FPGA and do not consume dedicated FPGA I/O ports.

\begin{figure*}[t]
\begin{center}
\includegraphics[width=.95\textwidth]{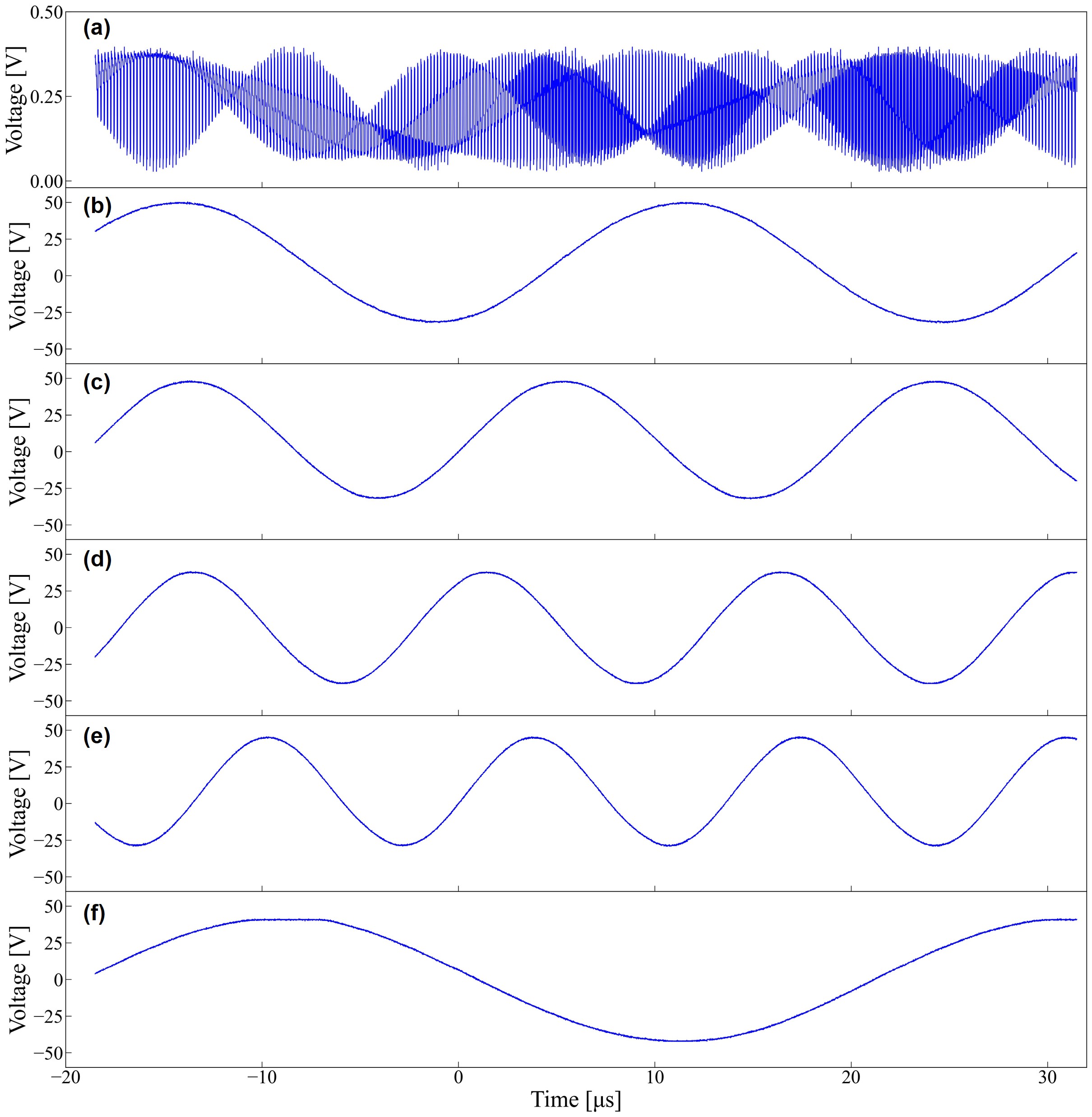}
\end{center}
\caption{
(a) Time-multiplexed signal comprising five sinusoidal waves, and (b--f) the corresponding output signals from Channels 0--4.
}
\label{fig:multiplexed}
\end{figure*}

\section{
System Characterization
}
\label{sec:poc}

Based on the concept proposed in Sec.~\ref{sec:general_concept}, we developed a proof-of-concept (PoC) system. 
Herein, we elucidate the PoC system and present the results of our evaluation of its analog output signals.

\subsection{
Development of PoC System
}

Figure~\ref{fig:design}(a) illustrates the developed system, which is capable of multiplexing five output channels. 
The system employs a 14-bit high-speed DAC (AD9707), whose output signal is amplified using an OP-amp (AD8099) before being distributed to each channel. 
Each channel is equipped with a high-speed switch (SN74AUC1G66DCKR), 30-pF charging capacitor, and subsequent OP-amp (ADA4700-1). 
This facilitates the system in outputting signals ranging from --50 to +50~V.

Notably, the developed board does not include a dedicated digital decoder. 
In contrast, demultiplexing is implemented through the direct routing of digital signals to each switch. 
Although this approach requires $N$ digital signals for a multiplexing factor of $N$, it is sufficient for this PoC.

The high-speed DAC on the board was controlled using an AMD ZCU106 Evaluation Board~\cite{zcu106}. 
The interface between the FPGA and DAC was implemented as a 14-bit parallel data interface comprising 14 data lines and 2 differential clock lines, all generated via the FPGA. 
The high-speed DAC operated at 30~Msps, resulting in a voltage update rate of 6~Msps for each output channel.

\subsection{Operational Tests}

We conducted operational tests using the developed system to assess its signal output performance. 
As an example, we generated a multiplexed signal composed of five sinusoidal waves. 
Figure~\ref{fig:multiplexed} shows both the time-multiplexed signal and the corresponding outputs from each channel: Fig.~\ref{fig:multiplexed}(a) shows the multiplexed signal, while Figs.~\ref{fig:multiplexed}(b) through \ref{fig:multiplexed}(f) display the sinusoidal outputs from each channel.

The multiplexed signal shown in Fig.~\ref{fig:multiplexed}(a) was acquired by connecting a coaxial cable with a 270~$\Omega$ resistor to the output of AD8099 and measuring it with an oscilloscope set to a 50~$\Omega$ input impedance. 
Consequently, the amplitude is reduced by a voltage divider factor of $50/(270+50)\approx0.156$ relative to the original output voltage. 
Note that the phases of the multiplexed signal and individual channel outputs are not aligned because each signal is measured independently. 
The results demonstrate that individual sinusoidal waveforms can be extracted successfully from the multiplexed signal.

\subsection{
Evaluation of the Settling Time of the High-speed DAC and the Capacitor Discharge-induced Voltage Drop
}

We measured the settling time of the high-speed DAC by evaluating the rise and fall times when varying output voltages between its minimum and maximum code values. 
The measured settling time was approximately 20~ns compared with the 11~ns specified in the AD9707 datasheet~\cite{ad9707}. 
This discrepancy may be attributed to the limited slew rate of the subsequent OP-amp, which extended the overall settling time.

\begin{figure}[t]
\begin{center}
\includegraphics[width=.48\textwidth]{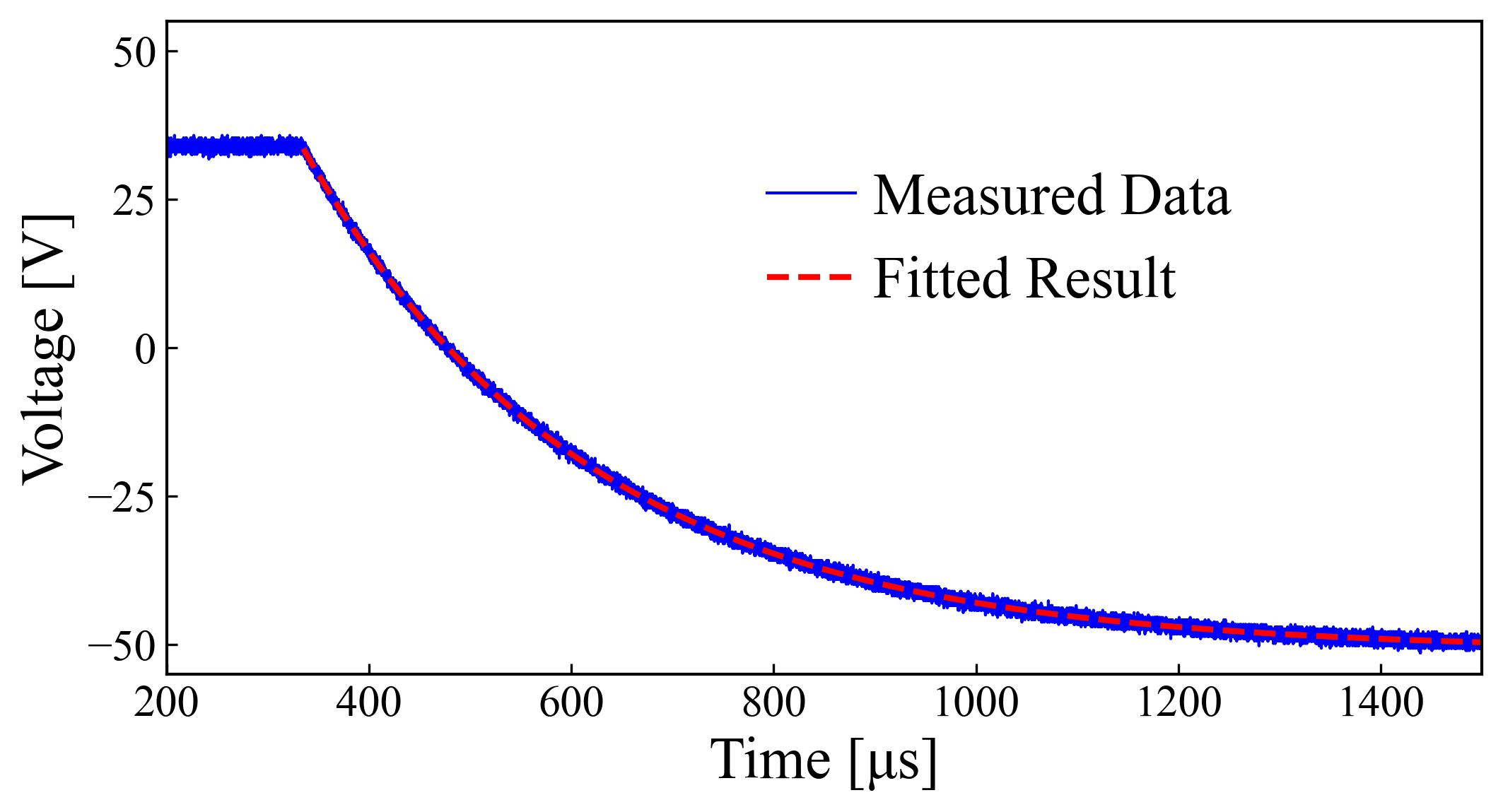}
\end{center}
\caption{
Measured capacitor discharge behavior in Channel 0. 
The red dashed curve indicates the exponential decay with a discharge constant of 282.6~$\mu$s.
}
\label{fig:discharge}
\end{figure}

The capacitor discharge behavior was evaluated. 
After charging the capacitor in Channel 0, the switch was released and its discharge was observed. 
The observed discharge of the capacitor is shown in Fig.~\ref{fig:discharge}.
We obtained a discharge time constant of 282.6~$\mu$s by fitting the data to an exponential decay model.

In our experiment, the high-speed DAC operated at 30~Msps, thereby allocating approximately 33.3~ns to each channel. 
Because we are multiplexing five output channels, the recharge cycle occurs approximately every 166.6~ns. 
Within this interval, the voltage drop was approximately 0.06\% of the initial voltage.

To further extend the discharge time constant, one could increase the capacitance.
However, this would result in a longer charging time, resulting in a trade-off that must be carefully balanced. 
Alternatively, the voltage drop can be controlled using a buffer or an OP-amp with a higher input impedance in the final stage.

\subsection{
Evaluation of Crosstalk Between Channels
}

Finally, we evaluated the interchannel crosstalk. 
In this experiment, a sinusoidal wave was output from Channel 4, and the signals from the other channels were measured. 
To mitigate the high-frequency noise likely originating from the switching process, a fifth-order Butterworth low-pass filter with a cutoff frequency of 70~kHz was applied to the outputs of each channel before analysis. 
The crosstalk was quantified by comparing the amplitude of each channel with that of Channel 4.

Table~\ref{tab:crosstalk} summarizes the crosstalk measurements obtained in these experiments. 
In all channels, crosstalk was below 60~dB. 
The crosstalk between channels 0 and 4 was slightly higher than that observed in the other channels. 
This could be due to the circuit configuration, in which Channel 0 is the first channel to be charged immediately after Channel 4, causing Channel 0's output to be influenced by the DAC code value of Channel 4.

\begin{table}[htbp]
\caption{Measured Crosstalk between Channel~4 and Channels~0--3}
\begin{center}
\begin{tabular}{|c|c|}
\hline
\textbf{Channel} & \textbf{Crosstalk [dB]} \\
\hline
Channel~0 & --62.8 \\
\hline
Channel~1 & --67.6 \\
\hline
Channel~2 & --67.7 \\
\hline
Channel~3 & --67.2 \\
\hline
\end{tabular}
\label{tab:crosstalk}
\end{center}
\end{table}

\section{
Conclusion and future work
}

This study proposed a scalable control scheme for large-scale QCCD trapped-ion quantum processors by leveraging a high-speed DAC. 
By generating complete voltage waveforms with a single high-speed DAC and sharing them among multiple electrodes, the proposed method significantly reduced the wiring complexity and overall resource requirements compared to conventional methods, which require one dedicated DAC per electrode. 
Based on realistic parameters and commercially available components, our analysis demonstrated that a QCCD system with 10,000 trap electrodes was controllable using only 13 FPGAs and 104 high-speed DACs. 
In addition, we developed a PoC electronic system based on this approach and evaluated its analog output performance. 
Consequently, we confirmed the ability of the system to effectively generate the voltage waveforms required for trap electrode control.

In future research, we will focus on the following objectives. 
We will integrate our control architecture with room-temperature trapped-ion setups. 
Further, we will optimize the digital and analog subsystems of the control architecture to improve resource efficiency and overall system robustness. 
Although this study primarily addresses ion shuttling and trapping, in principle, our control method is extendable to other operations required for QCCD architectures, such as split/combine~\cite{pino2021demonstration, home2004electrode, ruster2014experimental, kaufmann2014dynamics, kaushal2020shuttling, moses2023race, fallek2024rapid} and physical swap~\cite{pino2021demonstration, splatt2009deterministic, kaufmann2017fast, van2020coherent, kaushal2020shuttling, moses2023race}. 
These operations can be implemented by appropriately scheduling the electrode control sequences. 
Finally, considering the increased interest in the operation of trapped-ion quantum computers under cryogenic conditions~\cite{labaziewicz2008suppression, chiaverini2014insensitivity, brandl2016cryogenic, pagano2018cryogenic, weber2023cryogenic}, we aim to investigate the feasibility of the proposed scheme at low temperatures.

\bibliographystyle{IEEEtran}
\bibliography{ref}

\end{document}